\documentclass[aps,preprintnumbers, amsmath, amssymb]{revtex4}
\usepackage{braket} 
\usepackage{float}
\usepackage{graphics,epsfig}
\usepackage{graphicx}
\usepackage{dcolumn}
\usepackage{bm}
\usepackage{subfigure}
\usepackage{mathtools}
\begin{document}
\baselineskip=0.8 cm

\title{{\bf Entanglement dynamics of accelerated atoms with environment-induced interactions }}
\author{Xuerui Zhou, Chenhao Ma, Zixu Zhao\footnote{Corresponding author. zhao$_{-}$zixu@yeah.net}}
\affiliation{School of Science, Xi'an University of Posts and Telecommunications, Xi'an 710121, China}
\vspace*{0.2cm}
\begin{abstract}
\baselineskip=0.6 cm
\begin{center}
		{\bf Abstract}
\end{center}	
	
We investigate the influence of environment-induced interactions on the entanglement dynamics of two uniformly accelerated atoms coupled to a fluctuating massless scalar field with a reflecting boundary. The two atoms are aligned perpendicular to the boundary. The entanglement behavior is influenced by the competition between the environment-induced interatomic and the environment-induced atom-plate interactions, which can be characterized by certain critical values. The maximum of concurrence generated during evolution decreases non-monotonically with the acceleration, which implies that the anti-Unruh phenomenon can exist for some situations even when both environment-induced interatomic and environment-induced atom-plate interactions are considered.

\end{abstract}
\maketitle
\newpage
\vspace*{0.2cm}
\section{Introduction}

In 1935, Hermann studied the mathematics of an electron interacting with a photon and showed that, depending on the observational context, different wave functions can be obtained for the same system at the same instant \cite{Hermann1935}. Later that same year, Einstein, Podolsky and Rosen (EPR) introduced a thought experiment, which proposed that the wave function does not provide a complete description of the physical reality \cite{Einstein1935}. Schr\"{o}dinger used ``Verschr\"{a}nkung" to describe situations like that of the EPR scenario \cite{Schrodinger193501}. In 1946, Wheeler suggested an experiment to verify that the two quanta emitted in the annihilation of a positron-electron pair with zero relative angular momentum are polarized at right angles to each other \cite{Wheeler1946}. Wu and Shaknov observed the angular correlation of scattered annihilation radiation \cite{Wu1950}. Bohm proposed an interpretation of the quantum theory based on ``hidden" variables \cite{Bohm1952}. In 1964, Bell derived the inequality and showed that the statistical predictions of quantum mechanics are incompatible with separable predetermination \cite{Bell1964}. In 1972, Freedman and Clauser provided evidence against local hidden variable theories \cite{Freedman1972}. In 1982, Aspect et al. showed the violation of Bell's inequalities \cite{1Aspect1982,2Aspect1982}. Wootters and Zurek proposed that a single quantum cannot be cloned \cite{Zurek1982}. Bennett et al. showed a theoretical scheme for realizing quantum teleportation via dual classical and Einstein-Podolsky-Rosen channels \cite{Bennett1993}. In 1997, quantum teleportation was experimentally demonstrated \cite{Zeilinger1997}.

The entanglement dynamics of noninertial atoms is an interesting subject. A uniformly accelerated detector perceives the Minkowski vacuum as a thermal bath at a temperature proportional to its proper acceleration, which is the Unruh effect \cite{Fulling1973,Hawking1974,Hawking1975,Davies1975,Unruh1976}. Braun showed that entanglement can be generated between two qubits that interact with a common heat bath in thermal equilibrium, but have no direct interaction with each other \cite{Braun2002}. Kim et al. demonstrated that two atoms can become entangled through their interaction with a highly chaotic system depending on the initial preparation of the atoms \cite{Kim2002}. Benatti et al. showed that two noninteracting two-level systems in a common thermal bath can become mutually entangled via Markovian completely positive reduced evolution \cite{Benatti2003}, and that a uniformly accelerating system composed of two independent two-level atoms reaches the corresponding asymptotic equilibrium state that turns out to be entangled \cite{Benatti2004}.

The presence of a reflecting boundary can give rise to novel features. Zhang and Yu found that the presence of a boundary plays an important role in entanglement generation under certain circumstances and provides more freedom in manipulating entanglement generation \cite{J. Zhang2007}. Zhou et al. investigated the influence of boundaries on the quantum entanglement between local objects and a quantum field \cite{Zhou2013}. Ibort et al. showed that manipulating boundary conditions for a class of bipartite systems allows a separable state to evolve into an entangled one \cite{Ibort2014}. Cheng, Yu and Hu showed that the boundary greatly enriches the entanglement dynamics \cite{Cheng2018}. Cong, Tjoa and Mann performed the investigation of entanglement harvesting with moving mirrors \cite{Mann2019}.

According to the Gorini-Kossakowski-Lindblad-Sudarshan master equation \cite{Gorini1976,Lindblad1976,Breure2002} for open quantum systems, the environment induces decoherence, dissipation and an energy shift. The energy shift term contains both an environment-induced interatomic interaction and individual Lamb shifts. The influence of environment-induced interatomic interaction on the entanglement dynamics of a two-atom system was studied in Refs. \cite{chen2022,Liu2024,Ma2025}. On the other hand, the presence of a boundary renders the environment-induced Lamb shift position-dependent. Therefore, the environment-induced interactions include both an induced atom-atom interaction and an atom-boundary contribution. Recently, Chen, Hu and Yu investigated the influence of environment-induced interactions on the entanglement dynamics of two static atoms placed near a reflecting boundary \cite{Chen2026}. In this paper, we study  how environment-induced interactions affect the entanglement dynamics of two uniformly accelerated atoms aligned vertically with a reflecting boundary.

The structure of this paper is as follows. In Sec. II, we present the fundamental formulas for two uniformly accelerated two-level atoms coupled to vacuum scalar fields. In Sec. III, we will consider the effect of environment-induced interactions on the entanglement dynamics of accelerated atoms aligned perpendicularly to a reflecting boundary. We summarize our main results in the last section. We employ the natural units $\hbar = c = 1$ for convenience.

\section{The Basic Formulas}
We consider the interaction between two uniformly accelerated two-level atoms weakly coupled to a fluctuating massless scalar field in Minkowski vacuum with a perfectly reflecting boundary. The total Hamiltonian of the system is given by $H=H_{S}+H_{F}+H_{I}$. The Hamiltonian of the two-atom system is
\begin{equation}\label{hs}
	H_{S}=\frac{\omega}{2}\sigma_{3}^{(1)}+\frac{\omega}{2}\sigma_{3}^{(2)},
\end{equation}
where $\sigma^{(1)}_{i}=\sigma_{i}\otimes\sigma_{0}$ and $\sigma^{(2)}_{i}=\sigma_{0}\otimes\sigma_{i}$ with $\sigma_{i}~(i=1,2,3)$ standing for the Pauli matrices corresponding to atom $1$ and $2$, and $\sigma_{0}$ being the $2\times2$ unit matrix. $\omega$ is the energy level spacing of the atoms. $H_{F}$ represents the free Hamiltonian of the fluctuating scalar field. The interaction Hamiltonian $H_{I}$ can be written as \cite{Audretsch1994}
\begin{equation}\label{hi}
	H_{I}=\mu[\sigma^{(1)}_{2}\Phi(t,x_{1})+\sigma^{(2)}_{2}\Phi(t,x_{2})],
\end{equation}
where $\mu$ denotes the small coupling constant. We assume an initial state $\rho_{\mathrm{tot}}(0) = \rho(0) \otimes |0\rangle\langle 0|$, where $|0\rangle$ denotes the vacuum state of the field and $\rho(0)$ is the initial two-atom state. In the weak-coupling limit, the reduced dynamics of the two-atom system can be described by a Gorini-Kossakowski-Lindblad-Sudarshan master equation \cite{Gorini1976,Lindblad1976,Breure2002}
\begin{equation}\label{master1}
	\frac{\partial\rho(\tau)}{\partial\tau}=-i[H_{\rm eff},\rho(\tau)]+
	\mathcal{L}[\rho(\tau)],
\end{equation}
where the effective Hamiltonian is
\begin{equation}\label{master2}
	H_{\rm eff}=H_{S}-\frac{i}{2}\sum^{2}_{\alpha,\beta=1}\sum^{3}_{i,j=1}
	H_{ij}^{(\alpha\beta)}\sigma_{i}^{(\alpha)}\sigma_{j}^{(\beta)},
\end{equation}
and the dissipator is given by
\begin{equation}\label{master3}
	\mathcal{L}[\rho(\tau)]=\frac{1}{2}\sum^{2}_{\alpha,\beta=1}\sum^{3}_{i,j=1}
C_{ij}^{(\alpha\beta)}[2\sigma_{j}^{(\beta)}\rho\sigma_{i}^{(\alpha)}-
\sigma_{i}^{(\alpha)}\rho\sigma_{j}^{(\beta)}-\rho\sigma_{i}^{(\alpha)}
\sigma_{j}^{(\beta)}].
\end{equation}
Decoherence and dissipation arising from the environment are described by the dissipator $\mathcal{L}[\rho(\tau)]$. The coefficients $C_{ij}^{(\alpha\beta)}$ in the dissipator (\ref{master3}) take the form
\begin{equation}\label{cij}
	C_{ij}^{(\alpha\beta)}=A^{(\alpha\beta)}\delta_{ij}-iB^{(\alpha\beta)}
	\epsilon_{ijk}\delta_{3k}-A^{(\alpha\beta)}\delta_{3i}\delta_{3j},
\end{equation}
where
\begin{align}\label{AB}
	A^{(\alpha\beta)}=\frac{\mu^{2}}{4}[\mathcal{G}^{(\alpha\beta)}(\omega)+\mathcal{G}^{(\alpha\beta)}(-\omega)],\nonumber\\
	B^{(\alpha\beta)}=\frac{\mu^{2}}{4}[\mathcal{G}^{(\alpha\beta)}(\omega)-\mathcal{G}^{(\alpha\beta)}(-\omega)].
\end{align}
The Fourier transform $\mathcal{G}^{(\alpha\beta)}(\lambda)$ and the Hilbert transform $\mathcal{K}^{(\alpha\beta)}(\lambda)$ of the scalar field correlation function $\langle\Phi(\tau,x_{\alpha})\Phi(\tau',x_{\beta})\rangle$ determine the coefficients for the matrices $C_{ij}^{(\alpha\beta)}$ and $H_{ij}^{(\alpha\beta)}$. The Fourier transform is given by
\begin{equation}\label{gk}
	\mathcal{G}^{(\alpha\beta)}(\lambda)=\int^{\infty}_{-\infty}d\Delta\tau
	e^{i\lambda\Delta\tau}\langle\Phi(\tau,x_{\alpha})\Phi(\tau',x_{\beta})\rangle,
\end{equation}
and the Hilbert transform is
\begin{align}\label{gwhl}
	\mathcal{K}^{(\alpha\beta)}(\lambda)=\frac{P}{\pi i}\int^{\infty}_{-\infty}d\omega
	\frac{\mathcal{G}^{(\alpha\beta)}(\omega)}{\omega-\lambda},
\end{align}
where $P$ is the principal value. Similarly, $H_{ij}^{(\alpha\beta)}$ is obtained by substituting the Hilbert transform $\mathcal{K}^{(\alpha\beta)}(\lambda)$ for the Fourier transform $\mathcal{G}^{(\alpha\beta)}(\lambda)$. Then, the effective Hamiltonian is ${ H_{\rm eff}}=\tilde{H_{s}}+{ H^{(12)}_{\rm eff}}$. $\tilde{H_{s}}$ is a renormalization of the atomic Hamiltonian. It takes the same form as Eq. (\ref{hs}), but redefines the energy level spacing
\begin{equation}\label{omeg}
	\tilde{\omega}_\alpha = \omega - \frac{i\mu^2}{2}[\mathcal{K}^{(\alpha\alpha)}(\omega) - \mathcal{K}^{(\alpha\alpha)}(-\omega)].    \quad (\alpha = 1, 2)
\end{equation}
	${ H^{(12)}_{\rm eff}}$ represents the environment-induced coupling between the two atoms, and is given by
\begin{equation}\label{h12}
	H^{(12)}_{\rm eff}=-\sum_{i,j=1}^{3}\Omega_{ij}^{(12)}(\sigma_{i}\otimes\sigma_{j}),
\end{equation}
    where
\begin{equation}\label{omegaij}
	 \Omega_{ij}^{(12)}=\frac{i\mu^2}{4}\{[\mathcal{K}^{(12)}(\omega)+\mathcal{K}^{(12)}(-\omega)]\delta_{ij}-[\mathcal{K}^{(12)}(\omega)+\mathcal{K}^{(12)}(-\omega)]\delta_{3i}\delta_{3j}\}.
\end{equation}
Accordingly, the master equation takes the form
\begin{align}\label{master}
	\nonumber
	\frac{\partial \rho(\tau)}{\partial \tau}
    &= -i \sum_{\alpha=1}^{2} \tilde{\omega}_{\alpha}[ \sigma_{3}^{(\alpha)}, \rho(\tau)]
    + i \sum_{i,j=1}^{3} \Omega_{ij}^{(12)}[ \sigma_i \otimes \sigma_j, \rho(\tau)] \\
    &\quad + \frac{1}{2} \sum_{\alpha,\beta=1}^{2} \sum_{i,j=1}^{3} C_{ij}^{(\alpha\beta)}
    [ 2\sigma_{j}^{(\beta)} \rho \sigma_{i}^{(\alpha)}
    - \sigma_{i}^{(\alpha)} \rho \sigma_{j}^{(\beta)}
    - \rho \sigma_{i}^{(\alpha)} \sigma_{j}^{(\beta)}].
\end{align}

\section{Entanglement dynamics for accelerated atoms with a boundary}
\begin{figure}[htbp]
	\begin{center}	
		\includegraphics[scale=0.7]{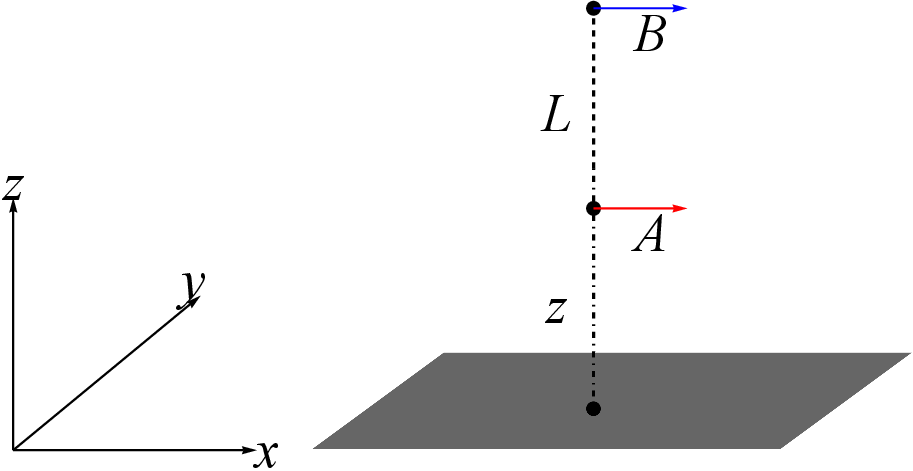}	
		\caption{\label{trajectory} The two accelerated atoms are separated with a distance $L$ and the reflecting boundary is located at $z=0$.}
	\end{center}
\end{figure}

In this section, we investigate the entanglement dynamics of two accelerated atoms aligned perpendicular to a reflecting boundary. Near a reflecting boundary, the environment-induced energy shifts of individual atoms are position-dependent. Therefore, the environment-induced interactions consist of atom-boundary contributions as well as the induced atom-atom interaction. As shown in Fig. \ref{trajectory}, both atoms experience the same uniform acceleration along the $x$ axis. $z$ is the distance from the boundary to the nearer atom, and $L$ is the distance between the two atoms. The trajectories of the two uniformly accelerated atoms are expressed as
\begin{align}\label{ab-tr}
	 &t_{1}(\tau)=\frac{1}{a} \sinh(a\tau),\;\;x_{1}(\tau)=\frac{1}{a} \cosh(a\tau),
\;\;y_{1}(\tau)=0,\;\;z_{1}(\tau)=z,\nonumber\\
&t_{2}(\tau)=\frac{1}{a} \sinh(a\tau),\;\;x_{2}(\tau)=\frac{1}{a} \cosh(a\tau),
\;\;y_{2}(\tau)=0,\;\;z_{2}(\tau)=z+L,
\end{align}
where $a$ is the proper acceleration. The Wightman function is given by \cite{Birrell1984}
\begin{align}\label{wigh-1}
	W\left(x, x'\right)=&-\frac{1}{4 \pi^{2}}[\frac{1}{(t-t'-i
	\epsilon)^{2}-(x-x')^{2}-(y-y')^{2}
	-(z-z')^{2}}\nonumber\\
	&-\frac{1}{(t-t'-i \epsilon)^{2}-(x-x')^{2}-(y-y')^{2}
		-(z+z')^{2}}]\;.
\end{align}
The Fourier transforms of the correlation functions take the form
\begin{align}\label{g11g12}
\nonumber
         \mathcal{G}^{(11)}(\omega)&=\frac{\omega}{4\pi} (\coth(\pi \omega / a)+1) [1 - f(\omega, z)],\\
\mathcal{G}^{(22)}(\omega)&=\frac{\omega}{4\pi} (\coth(\pi \omega / a)+1) [1 - f(\omega, L+z)],\nonumber\\
\mathcal{G}^{(12)}(\omega)&=\mathcal{G}^{(21)}(\omega)=\frac{\omega}{4\pi} (\coth(\pi \omega / a)+1) [ f(\omega, L/2)-f(\omega, z+L/2)],
\end{align}
where
\begin{align}\label{01}	
 f(\omega ,z)=\frac{\sin[\frac{2\omega}{a} \sinh^{-1}(a z)]}{2 z \omega \sqrt{1 + a^2 z^2}}.
\end{align}
Then, the coefficient matrices $C_{ij}^{(\alpha\beta)}$ and $\Omega_{ij}^{(12)}$ can be written as
\begin{eqnarray}\label{coefficient matrix1}
&&C^{(11)}_{ij}=A_1\delta_{ij}-iB_1\epsilon_{ijk}\delta_{3k}-A_1\delta_{3i}\delta_{3j},\nonumber\\
&&C^{(22)}_{ij}=A_2\delta_{ij}-iB_2\epsilon_{ijk}\delta_{3k}-A_2\delta_{3i}\delta_{3j},\nonumber\\
&&C^{(12)}_{ij}=A_3\delta_{ij}-iB_3\epsilon_{ijk}\delta_{3k}-A_3\delta_{3i}\delta_{3j},\nonumber\\
&&C^{(21)}_{ij}=A_4\delta_{ij}-iB_4\epsilon_{ijk}\delta_{3k}-A_4\delta_{3i}\delta_{3j},\nonumber\\
&&\Omega^{(12)}_{ij}=D\delta_{ij}-D\delta_{3i}\delta_{3j},
\end{eqnarray}
where
\begin{align*}
\nonumber
A_{1}&=\frac{\Gamma_0}{4}\coth(\pi \omega / a)[(1 - f(\omega, z)],\\
B_{1}&=\frac{\Gamma_0}{4}[1 - f(\omega, z)],\nonumber\\
A_{2}&=\frac{\Gamma_0}{4}\coth(\pi \omega / a)[1 - f(\omega, L+z)],\nonumber\\
B_{2}&=\frac{\Gamma_0}{4}[1 - f(\omega, L+z)],\nonumber\\
\end{align*}
\newpage
\begin{align}\label{A1A2B1B2D}
A_{3}&=A_{4}=\frac{\Gamma_0}{4}\coth(\pi \omega / a)[f(\omega, L/2) - f(\omega, z + L/2)],\nonumber\\
B_{3}&=B_{4}=\frac{\Gamma_0}{4}[f(\omega, L/2) - f(\omega, z + L/2)],\nonumber\\
{D}&=\frac{\Gamma_0}{4}[h(\omega, L/2) - h(\omega, z + L/2)],
\end{align}
with
\begin{align}\label{02}
h(\omega ,z)=\frac{\cos[\frac{2\omega}{a} \sinh^{-1}(a z)]}{2 z \omega \sqrt{1 + a^2 z^2}},
\end{align}
and $\Gamma_{0}=\frac{\mu^{2}\omega}{2\pi}$ is the spontaneous emission rate for an inertial atom in the Minkowski vacuum. We adopt the coupled basis $\{|G\rangle=|00\rangle, |A\rangle=\frac{1}{\sqrt{2}}(|10\rangle-|01\rangle), |S\rangle=\frac{1}{\sqrt{2}}(|10\rangle+|01\rangle), |E\rangle=|11\rangle\}$. Then, the time evolution equations of the density matrix elements are
\begin{eqnarray*}
\dot{\rho}_{GG}&=&-2(A_1+A_2-B_1-B_2)\rho_{GG}+(A_1+A_2-A_3-A_4+B_1+B_2-B_3-B_4)\rho_{AA}\nonumber\\&&+(A_1+A_2+A_3+A_4+B_1+B_2+B_3+B_4)\rho_{SS}+(A_1-A_2-A_3+A_4+B_1\nonumber\\&&-B_2-B_3+B_4)\rho_{AS}+(A_1-A_2+A_3-A_4+B_1-B_2+B_3-B_4)\rho_{SA},\\
\dot{\rho}_{EE}&=&-2(A_1+A_2+B_1+B_2)\rho_{EE} + (A_1+A_2-A_3-A_4-B_1-B_2+B_3+B_4)\rho_{AA}\nonumber\\&&+(A_1+A_2+A_3+A_4-B_1-B_2-B_3-B_4)\rho_{SS}+(-A_1+A_2+A_3-A_4+B_1\nonumber\\&&-B_2-B_3+B_4)\rho_{AS}+(-A_1+A_2-A_3+A_4+B_1-B_2+B_3-B_4)\rho_{SA},\\
\dot{\rho}_{AA}&=&-2(A_1+A_2-A_3-A_4)\rho_{AA} +(A_1+A_2-A_3-A_4-B_1-B_2+B_3+B_4) \rho_{GG},\nonumber\\&&+(A_1+A_2-A_3-A_4+B_1+B_2-B_3-B_4)\rho_{EE}+(-B_1+B_2+B_3-B_4)\rho_{AS}\nonumber\\&&+(-B_1+B_2+B_3-B_4)\rho_{SA}-i\Delta(\rho_{AS}-\rho_{SA}),\\
\dot{\rho}_{SS}&=&-2(A_1+A_2+A_3+A_4)\rho_{SS} +(A_1+A_2+A_3+A_4-B_1-B_2-B_3-B_4)
\rho_{GG}\nonumber\\&&+(A_1+A_2+A_3+A_4+B_1+B_2+B_3+B_4)\rho_{EE}+(-B_1+B_2+B_3-B_4)\rho_{AS}\nonumber\\&&+(-B_1+B_2+B_3-B_4)\rho_{SA}+i\Delta(\rho_{AS}-\rho_{SA}),\\
\dot{\rho}_{AS}&=&(A_1-A_2-A_3+A_4-B_1+B_2+B_3-B_4)\rho_{GG}+(-A_1+A_2+A_3-A_4-B_1\nonumber\\&&+B_2+B_3-B_4)\rho_{EE}+(-B_1+B_2+B_3-B_4)
\rho_{AA}+(-B_1+B_2+B_3-B_4)\rho_{SS}\nonumber\\&&-2(A_1+A_2+2iD)\rho_{AS}+i\Delta(\rho_{SS}-\rho_{AA}),
\end{eqnarray*}
\newpage
\begin{eqnarray}\label{pf26}
\dot{\rho}_{SA}&=&(A_1-A_2+A_3-A_4-B_1+B_2-B_3+B_4)\rho_{GG}+(-A_1+A_2-A_3+A_4-B_1\nonumber\\&&+B_2-B_3+B_4)\rho_{EE}+(-B_1+B_2+B_3-B_4)
\rho_{AA}+(-B_1+B_2+B_3-B_4)\rho_{SS}\nonumber\\&&-2(A_1+A_2-2iD)\rho_{SA}-i\Delta(\rho_{SS}-\rho_{AA}),\nonumber\\
\dot{\rho}_{GE}&=&-2(A_1+A_2)\rho_{GE}, \;\;\;\;\;\;\;\;\; \;\;\;\;\;\;\;\dot{\rho}_{EG}=-2(A_1+A_2)\rho_{EG},\label{state1}
\end{eqnarray}
where $\rho_{IJ}=\langle I|\rho|J\rangle,I,J\in\{G,E,A,S\}$, and
\begin{align}\label{03}
\Delta = \tilde{\omega}_2 - \tilde{\omega}_1= \frac{\Gamma_0}{2} \operatorname{coth}(\pi \omega/a) [h(\omega, z+L) - h(\omega, z)],
\end{align}
represents the difference in the environment-induced energy shifts of the two atoms. For simplicity, the initial state is chosen to be an X state, namely, the nonzero elements are arranged along the diagonal and antidiagonal of the density matrix. Eq. \eqref{pf26} shows that these elements evolve independently of the remaining components, thus preserving the X structure. The concurrence \cite{Wootters1998}, as a measurement of quantum entanglement for a two-atom system in the X state, is given by \cite{Ficek2004}
\begin{align}\label{pfc}
	C[\rho(\tau)]=\max\{0,K_{1}(\tau),K_{2}(\tau)\},
\end{align}
	where
\begin{align}\label{pfk}
	\nonumber K_{1}(\tau)=&\sqrt{[\rho_{AA}(\tau)-\rho_{SS}(\tau)]^{2}-[\rho_{AS}(\tau)-\rho_{SA}(\tau)]^{2}}-
	2\sqrt{\rho_{GG}(\tau)\rho_{EE}(\tau)},\\
	K_{2}(\tau)=&2|\rho_{GE}(\tau)|-\sqrt{[\rho_{AA}(\tau)+\rho_{SS}(\tau)]^{2}-
		[\rho_{AS}(\tau)+\rho_{SA}(\tau)]^{2}}.
\end{align}

From Eq. \eqref{pf26}, for atoms aligned perpendicular to the boundary, the environment-induced interactions influence the entanglement dynamics for any initial state, not only for a restricted class satisfying $\rho_{AS}(0) \neq 0$ and $\rho_{SA}(0) \neq 0$ as in free space. When both atoms are far from the boundary, the atom-boundary contribution becomes negligible.

In the following, we study the time evolution of entanglement for the separable state $\left| 10 \right\rangle$ and the maximally entangled antisymmetric state $\left| A \right\rangle$. We also analyze the maximum concurrence generated during the evolution with initial $\left| 10 \right\rangle$. For both initial states, it is clear that $\rho_{GE}(\tau) = \rho_{EG}(\tau) = 0$, so $K_{2}(\tau)<0$ and
\begin{align}\label{k11}
C[\rho(\tau)]= \max\{0,K_{1}(\tau)\}.
\end{align}

\subsection{Time evolution of entanglement}
We investigate the generation and degradation of entanglement in two-atom systems initially prepared in the $\left| 10 \right\rangle$ and $\left| A \right\rangle$ states, respectively.
\subsubsection{Entanglement dynamics for two-atom system with initial state $\left| 10 \right\rangle$}
We study entanglement generation near the initial time. For a two-atom system prepared in the initial state $\left| 10 \right\rangle$, since $K_{1}(0)=0$, entanglement can be generated near the initial time $\tau=0$ when $K_{1}'(0)>0$. We can obtain
\begin{align}\label{k11}
{K}_{1}'(0)=4\sqrt{A_3^2 + D^2} - 4\sqrt{(A_2 - B_2)(A_1 + B_1)}.
\end{align}
For entanglement to be generated near the initial time, the condition is
\begin{align}\label{pf8}
A_{3}^{2}+D^{2}>(A_2 - B_2)(A_1 + B_1),
\end{align}
which does not depend on $\Delta$. Hence, the difference in the individual environment-induced energy shifts does not influence the condition for early-time entanglement generation.

\begin{figure}[htbp]
	\begin{center}		
        \includegraphics[scale=0.7]{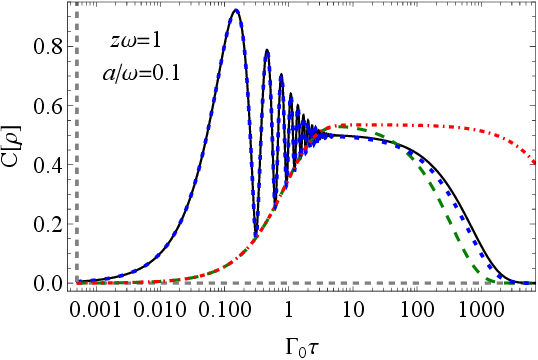}\vspace{0.01\textwidth}\hspace{0.02\textwidth}
	    \includegraphics[scale=0.7]{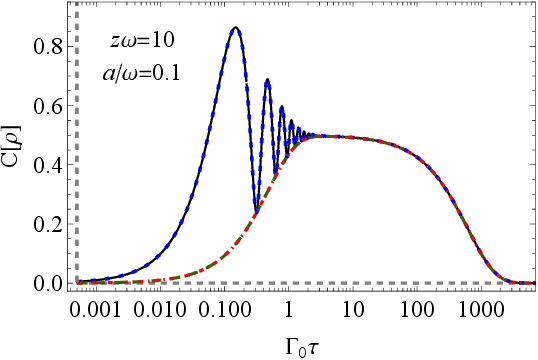}\vspace{0.01\textwidth}
        \includegraphics[scale=0.7]{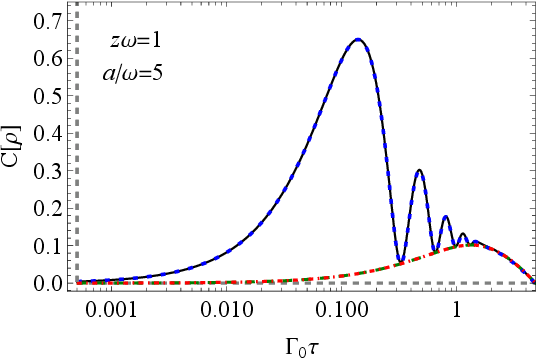}\vspace{0.01\textwidth}\hspace{0.02\textwidth}
	    \includegraphics[scale=0.7]{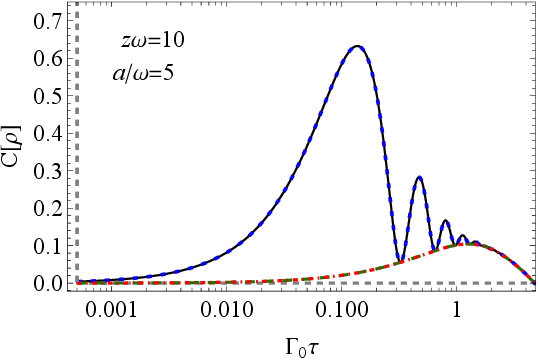}\vspace{0.01\textwidth}
\caption{\label{cpL0.1} The time evolution of concurrence for $L\omega=0.1$ with $z\omega=\{1, 10\}$ and $a/\omega=\{0.1, 5\}$ initially in the state $\left| 10 \right\rangle$. The solid (black) and dot-dashed (red) lines represent uniformly accelerated atoms with and without environment-induced interactions, respectively. The dotted line (blue) represents only the environment-induced atom-atom interaction, and the dashed line (green) represents only the environment-induced atom-plate interaction.}
    \end{center}
\end{figure}

In Fig. \ref{cpL0.1}, we depict the time evolution of concurrence for the two-atom system with a small interatomic separation $L\omega=0.1$ initially in the state $\left| 10 \right\rangle$. We compare the following four cases: (i) atoms with both environment-induced interactions ($D \neq 0,\, \Delta \neq 0$); (ii) without environment-induced interactions ($D=0,\, \Delta=0$); (iii) with only the environment-induced atom-plate interaction ($D=0,\, \Delta \neq 0$); (iv) with only the environment-induced interatomic interaction ($D \neq 0,\, \Delta = 0$). For a small $a/\omega={0.1}$ and a not large $z\omega=1$, the solid black line is higher than the dot-dashed red line because the blue dotted line dominates at early times, which indicates that the environment-induced interatomic interaction promotes the generation of entanglement. Later, the dashed green line may be higher or lower than the blue dotted line, but the combined effect suppresses the generation of entanglement. The environment-induced interactions can be ignored at later times for a large $a/\omega={5}$ or a large $z\omega=10$.

\begin{figure}[htbp]
	\begin{center}	
	    \includegraphics[scale=0.55]{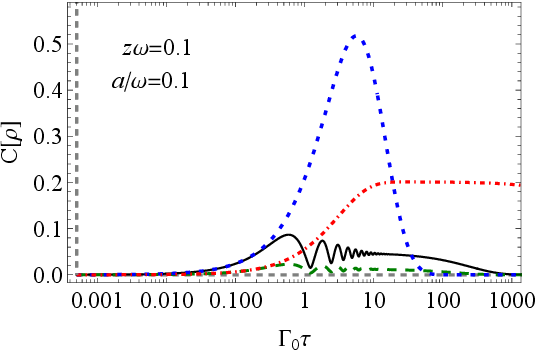}\vspace{0.01\textwidth}\hspace{0.02\textwidth}
		\includegraphics[scale=0.55]{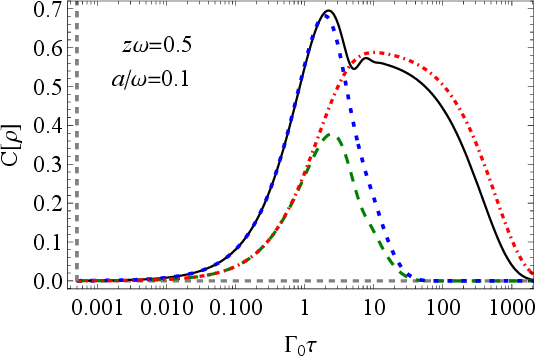}\vspace{0.01\textwidth}\hspace{0.02\textwidth}
		\includegraphics[scale=0.55]{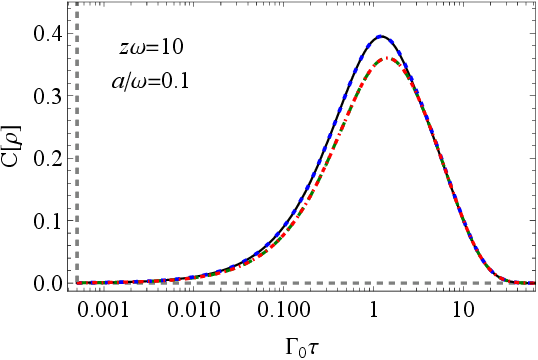}\vspace{0.01\textwidth}
		\includegraphics[scale=0.55]{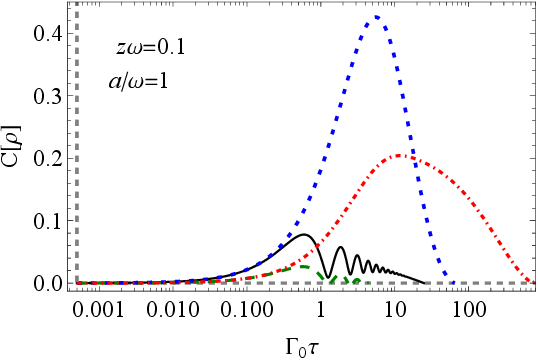}\vspace{0.01\textwidth}\hspace{0.02\textwidth}
		\includegraphics[scale=0.55]{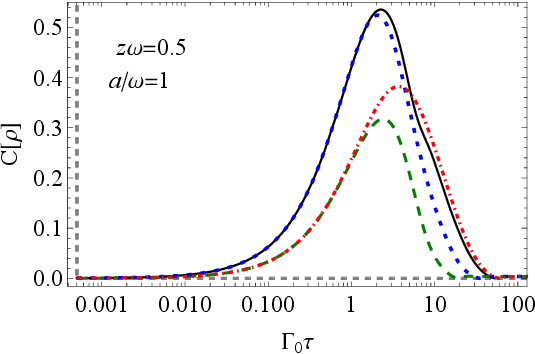}\vspace{0.01\textwidth}\hspace{0.02\textwidth}
		\includegraphics[scale=0.55]{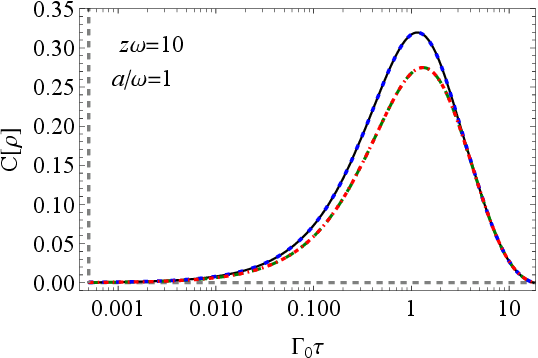}\vspace{0.01\textwidth}
\end{center}\caption{\label{cpL1}The time evolution of concurrence for $L\omega=1$ with $z\omega=\{0.1, 0.5, 10\}$ and $a/\omega=\{0.1, 1\}$ initially in the state $\left| 10 \right\rangle$. The solid (black) and dot-dashed (red) lines represent uniformly accelerated atoms with and without environment-induced interactions, respectively. The dotted line (blue) represents only the environment-induced atom-atom interaction, and the dashed line (green) represents only the environment-induced atom-plate interaction.}
\end{figure}

\begin{figure}[htbp]
	\begin{center}	
	    \includegraphics[scale=0.7]{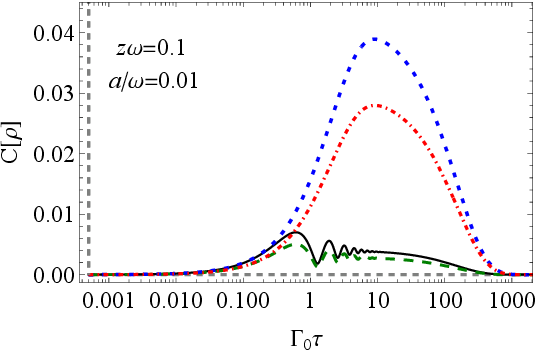}\vspace{0.01\textwidth}\hspace{0.02\textwidth}
		\includegraphics[scale=0.7]{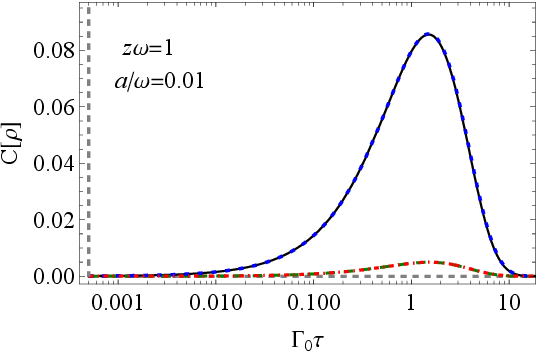}\vspace{0.01\textwidth}
		\includegraphics[scale=0.7]{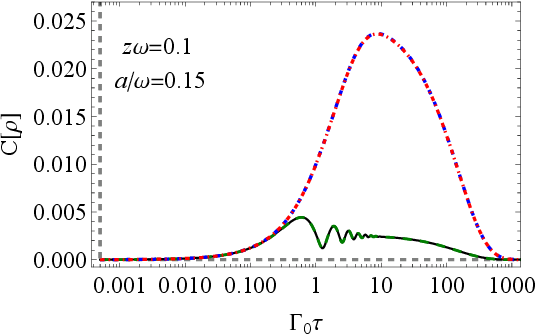}\vspace{0.01\textwidth}\hspace{0.02\textwidth}
		\includegraphics[scale=0.7]{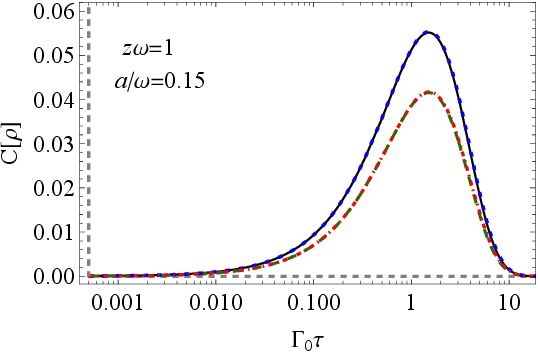}\vspace{0.01\textwidth}
        \includegraphics[scale=0.7]{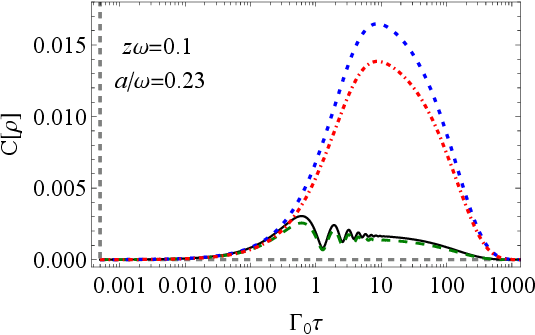}\vspace{0.01\textwidth}\hspace{0.02\textwidth}
		\includegraphics[scale=0.7]{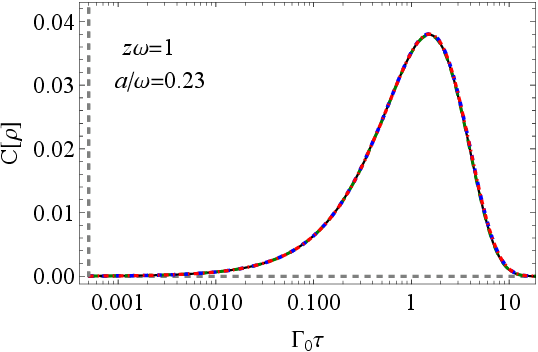}\vspace{0.01\textwidth}
\end{center}\caption{\label{cpL10}The time evolution of concurrence for $L\omega=10$ with $z\omega=\{0.1, 1\}$ and $a/\omega=\{0.01, 0.15, 0.23\}$ initially in the state $\left| 10 \right\rangle$. The solid (black) and dot-dashed (red) lines represent uniformly accelerated atoms with and without environment-induced interactions, respectively. The dotted line (blue) represents only the environment-induced atom-atom interaction, and the dashed line (green) represents only the environment-induced atom-plate interaction.}
\end{figure}

In Fig. \ref{cpL1}, we describe the time evolution of concurrence for the two-atom system with an intermediate interatomic separation $L\omega=1$ initially in the state $\left| 10 \right\rangle$.
Taking a small $a/\omega=0.1$, there exist two critical values $z_{c1}\omega\approx0.479$ and $z_{c2}\omega\approx0.965$, which lead to a transition in the entanglement behavior. For $z\omega=0.1<z_{c1}\omega$, the environment-induced interatomic interaction dominates at early times and enhances entanglement generation. Later, the environment-induced atom-plate interaction becomes dominant and inhibits entanglement generation. For $z\omega=0.5$, which satisfies $z_{c1}\omega<z\omega<z_{c2}\omega$, when (both) environment-induced interactions are considered, higher peaks of $C[\rho]$ can be obtained. For $z\omega=10>z_{c2}\omega$, the environment-induced interactions have little influence on the entanglement generation at later times. When we take a not small $a/\omega=1$, for a small $z\omega=0.1$, the entanglement is also suppressed at later times without the environment-induced interactions. For a not small $z\omega$, the behavior of $C[\rho]$ is similar to that for the small acceleration case.

As shown in Fig. \ref{cpL10}, we plot the time evolution of concurrence for the two-atom system with a large interatomic separation $L\omega=10$ initially in the state $\left| 10 \right\rangle$. Taking a small $z\omega=0.1$, there exist two critical values $a_{c1}/\omega\approx0.142$ and $a_{c2}/\omega\approx0.176$. When we consider $a/\omega=0.15$, which satisfies $a_{c1}/\omega<a/\omega<a_{c2}/\omega$, the environment-induced interactions have little effect on entanglement generation at early times. Later, the environment-induced atom-plate interaction becomes dominant and suppresses entanglement generation, whereas the environment-induced interatomic interaction can be ignored. When we take $a/\omega=0.01<a_{c1}/\omega$ or $a/\omega=0.23>a_{c2}/\omega$, $C[\rho]$ is enhanced at early times and suppressed at later times with environment-induced interactions considered. For a not small $z\omega=1$, the environment-induced interactions can be neglected at a particular acceleration.

\subsubsection{Entanglement dynamics for two-atom system with initial state $\left| A \right\rangle$}
We now investigate the entanglement degradation for two atoms initially prepared in the maximally entangled antisymmetric state $\left| A \right\rangle$.

\begin{figure}[htbp]
   \begin{center}
       \includegraphics[scale=0.7]{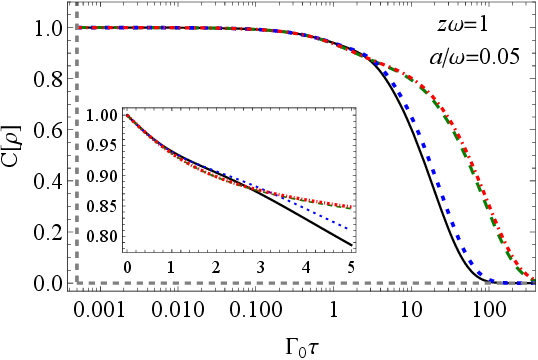}\hspace{0.02\textwidth}		
	   \includegraphics[scale=0.7]{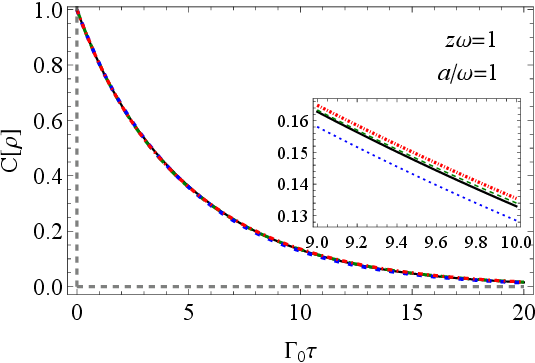}
\end{center}\caption{\label{1CP0} The time evolution of concurrence for $L\omega=1$ with $z\omega=1$ and $a/\omega=\{0.05, 1\}$ initially in the state $|A\rangle$. The solid (black) and dot-dashed (red) lines represent uniformly accelerated atoms with and without environment-induced interactions, respectively. The dotted line (blue) represents only the environment-induced atom-atom interaction, and the dashed line (green) represents only the environment-induced atom-plate interaction.}
\end{figure}

\begin{figure}[htbp]
   \begin{center}
        \includegraphics[scale=0.7]{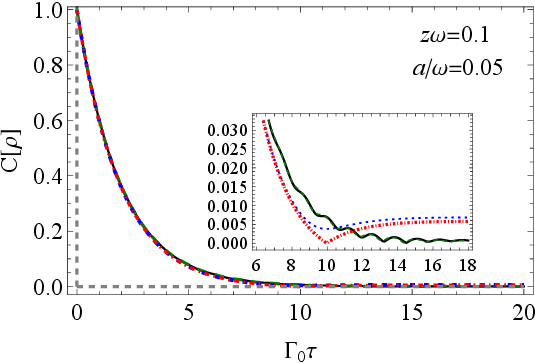}\vspace{0.01\textwidth}\hspace{0.02\textwidth}
	    \includegraphics[scale=0.7]{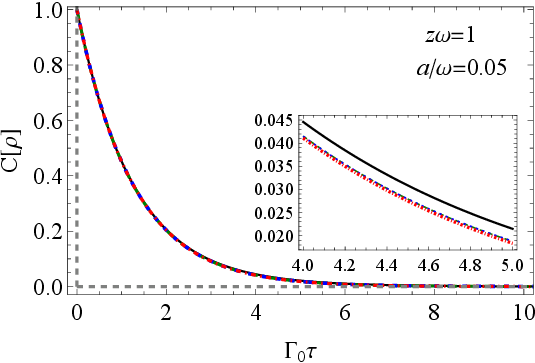}\vspace{0.01\textwidth}
        \includegraphics[scale=0.7]{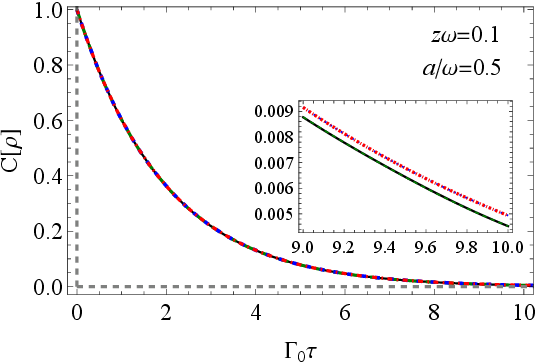}\hspace{0.02\textwidth}		
	    \includegraphics[scale=0.7]{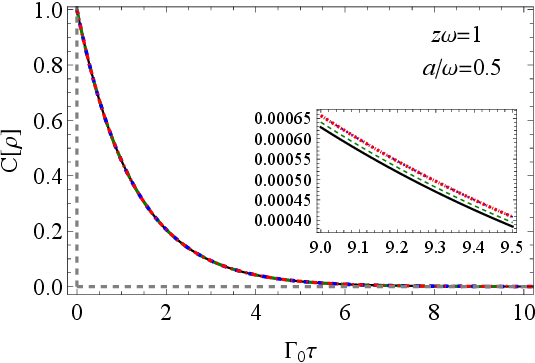}
	    \includegraphics[scale=0.7]{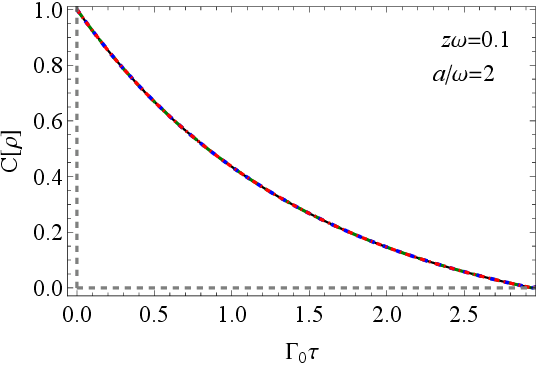}\hspace{0.02\textwidth}		
	    \includegraphics[scale=0.7]{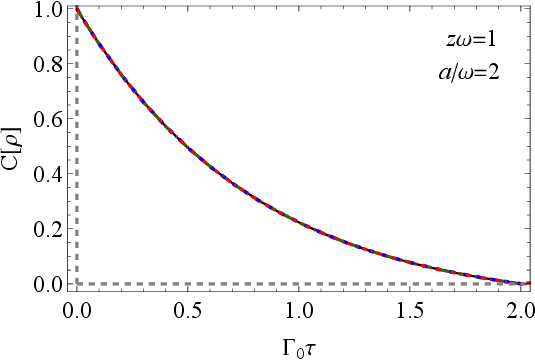}
\end{center}\caption{\label{2CP0} The time evolution of concurrence for $L\omega=20$ with $z\omega=\{0.1, 1\}$ and $a/\omega=\{0.05, 0.5, 2\}$ in the state $|A\rangle$. The solid (black) and dot-dashed (red) lines represent uniformly accelerated atoms with and without environment-induced interactions, respectively. The dotted line (blue) represents only the environment-induced atom-atom interaction, and the dashed line (green) represents only the environment-induced atom-plate interaction.}
\end{figure}

In Figs. \ref{1CP0} and \ref{2CP0}, we show the time evolution of entanglement for the two-atom system with the initial state $\left| A \right\rangle$. For a not large $L\omega=1$ and $z\omega=1$ in Fig. \ref{1CP0}, with a small $a/\omega=0.05$, the environment-induced interatomic interaction dominates at early times, and the decay of entanglement is slower than that without the environment-induced interactions. Later, both interactions accelerate the entanglement decay. For a not small $a/\omega=1$, the decay of entanglement is very rapid, but the environment-induced interactions have only a small effect on the decay. We take a large $L\omega=20$ in Fig. \ref{2CP0}. We first consider a small $z\omega=0.1$, where there exist two critical values $a_{c3}/\omega\approx0.476$ and $a_{c4}/\omega\approx1.835$. For $a/\omega=0.05<a_{c3}/\omega$, the environment-induced interactions have little influence on the early entanglement dynamics. Later, the environment-induced atom-plate interaction becomes dominant, first slowing the entanglement decay and then accelerating it. For $a/\omega=0.5$, which satisfies $a_{c3}/\omega<a/\omega<a_{c4}/\omega$, the environment-induced atom-plate interaction weakly accelerates the entanglement decay at later times, while the environment-induced interatomic interaction can be ignored. For $a/\omega=2>a_{c4}/\omega$, the environment-induced interactions have almost no impact on entanglement dynamics at any time. Taking a not small $z\omega=1$, there exist two critical values $a_{c5}/\omega\approx0.160$ and $a_{c6}/\omega\approx0.954$. For $a/\omega=0.05<a_{c5}/\omega$, the environment-induced interactions slow the decay of entanglement. For $a/\omega=0.5$, which satisfies $a_{c5}/\omega<a/\omega<a_{c6}/\omega$, the environment-induced atom-plate interaction also weakly accelerates the entanglement decay. For $a/\omega=2>a_{c6}/\omega$, the behavior of $C[\rho]$ is similar to the case with $z\omega=0.1$, which means that the large acceleration hides the effect of the environment-induced interactions.

\subsection{The maximum of concurrence generated during evolution}

\begin{figure}[htbp]
	\begin{center}		
    \includegraphics[scale=0.7]{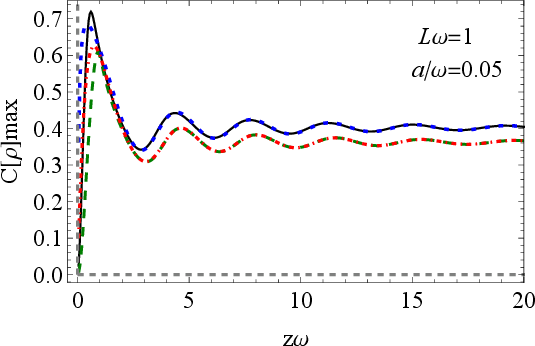}\vspace{0.01\textwidth}\hspace{0.02\textwidth}
	\includegraphics[scale=0.7]{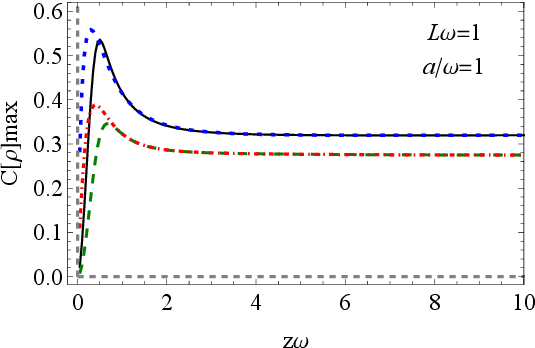}\vspace{0.01\textwidth}
\caption{\label{zwmax1} The maximum of concurrence $C[\rho]_{max}$ during evolution is plotted as a function of $z\omega$ with $L\omega=1$ and $a/\omega=\{0.05, 1\}$ initially in the state $\left| 10 \right\rangle$. The solid (black) and dot-dashed (red) lines represent uniformly accelerated atoms with and without environment-induced interactions, respectively. The dotted line (blue) represents only the environment-induced atom-atom interaction, and the dashed line (green) represents only the environment-induced atom-plate interaction.}
    \end{center}
\end{figure}

\begin{figure}[htbp]
	\begin{center}		
    \includegraphics[scale=0.7]{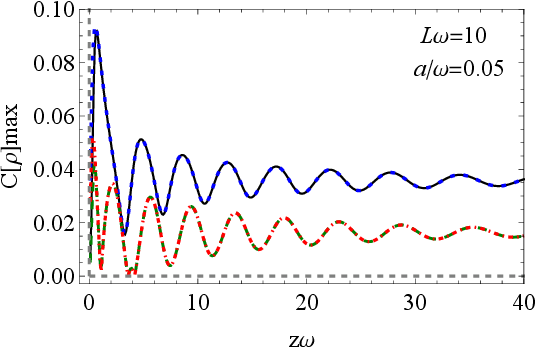}\vspace{0.01\textwidth}\hspace{0.02\textwidth}
	\includegraphics[scale=0.7]{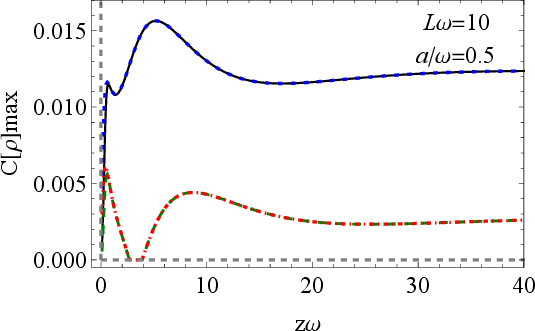}\vspace{0.01\textwidth}
\caption{\label{zwmax2} The maximum of concurrence $C[\rho]_{max}$ during evolution is plotted as a function of $z\omega$ with $L\omega=10$ and $a/\omega=\{0.05, 0.5\}$ initially in the state $\left| 10 \right\rangle$. The solid (black) and dot-dashed (red) lines represent uniformly accelerated atoms with and without environment-induced interactions, respectively. The dotted line (blue) represents only the environment-induced atom-atom interaction, and the dashed line (green) represents only the environment-induced atom-plate interaction.}
    \end{center}
\end{figure}

In Figs. \ref{zwmax1} and \ref{zwmax2}, we plot the maximum of concurrence $C[\rho]_{max}$ generated during evolution as a function of the distance between the nearer atom and the plate initially prepared in the state $\left| 10 \right\rangle$. For a not large $L\omega=1$ in Fig. \ref{zwmax1}, with a small $a/\omega=0.05$, $C[\rho]_{max}$ first increases to a maximum and then oscillates, approaching a steady value as $z\omega$ increases. We observe that $C[\rho]_{max}$ is dominated by the environment-induced atom-plate interaction for small $z\omega$, whereas the environment-induced interatomic interaction becomes dominant as $z\omega$ increases. The oscillatory behavior disappears for a not small $a/\omega=1$. Taking a large $L\omega=10$ in Fig. \ref{zwmax2}, when the environment-induced interactions are considered, for a small $a/\omega=0.05$, the oscillatory behavior of $C[\rho]_{max}$ persists in the large $z\omega$ regime. For a not small $a/\omega=0.5$, $C[\rho]_{max}$ varies non-monotonically at not small $z\omega$ and approaches a steady value as $z\omega$ increases. When the environment-induced interactions are neglected, $C[\rho]_{max}$ is close to zero for certain values of $z\omega$.

\begin{figure}[htbp]
	\begin{center}	
	    \includegraphics[scale=0.55]{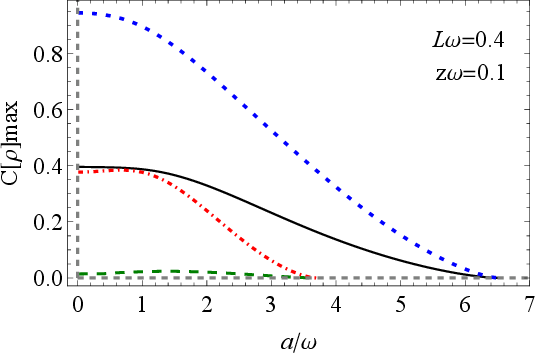}\vspace{0.01\textwidth}\hspace{0.02\textwidth}
		\includegraphics[scale=0.55]{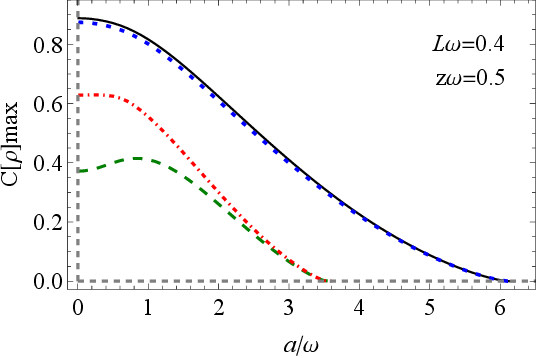}\vspace{0.01\textwidth}\hspace{0.02\textwidth}
		\includegraphics[scale=0.55]{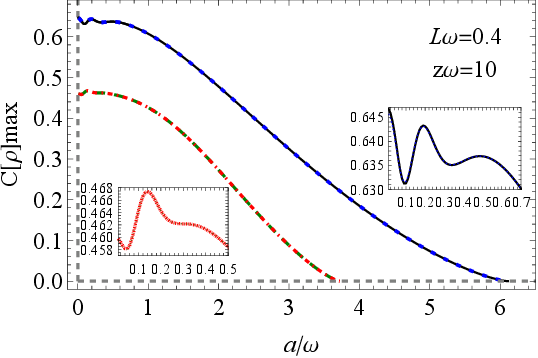}\vspace{0.01\textwidth}
        \includegraphics[scale=0.55]{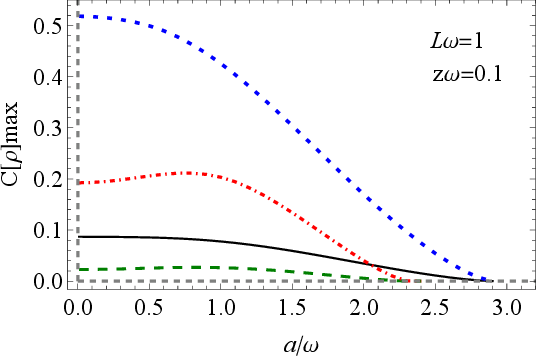}\vspace{0.01\textwidth}\hspace{0.02\textwidth}
		\includegraphics[scale=0.55]{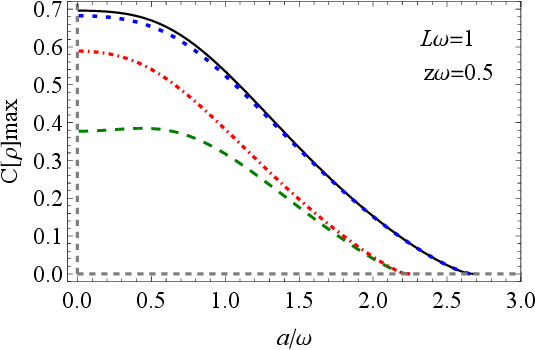}\vspace{0.01\textwidth}\hspace{0.02\textwidth}
		\includegraphics[scale=0.55]{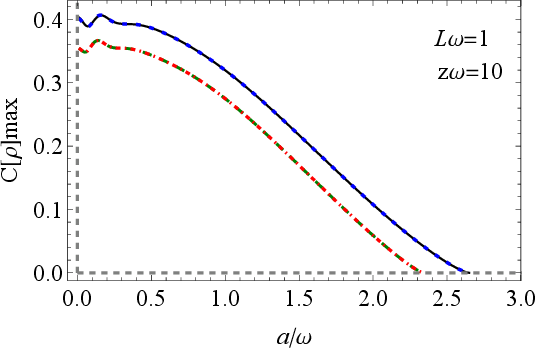}\vspace{0.01\textwidth}
		\includegraphics[scale=0.55]{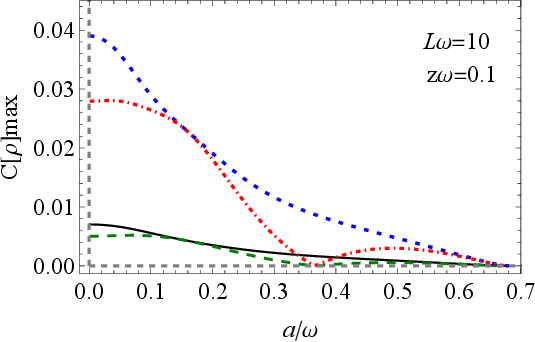}\vspace{0.01\textwidth}\hspace{0.02\textwidth}
		\includegraphics[scale=0.55]{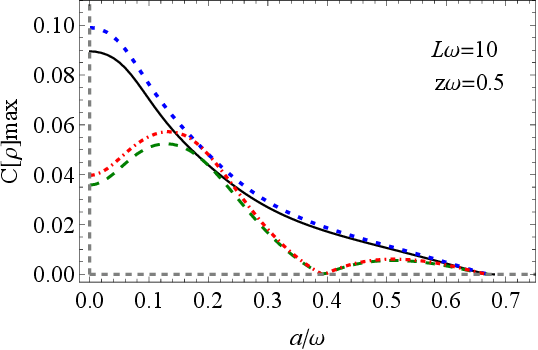}\vspace{0.01\textwidth}\hspace{0.02\textwidth}
		\includegraphics[scale=0.55]{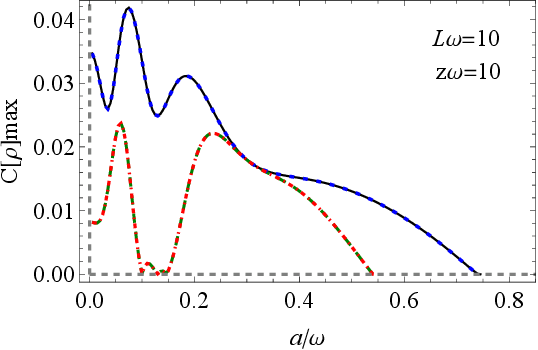}\vspace{0.01\textwidth}
\end{center}\caption{\label{awmax} The maximum of concurrence $C[\rho]_{max}$ during evolution is plotted as a function of $a/\omega$ with $L\omega=\{0.4, 1, 10\}$ and $z\omega=\{0.1, 0.5, 10\}$ initially in the state $\left| 10 \right\rangle$. The solid (black) and dot-dashed (red) lines represent uniformly accelerated atoms with and without environment-induced interactions, respectively. The dotted line (blue) represents only the environment-induced atom-atom interaction, and the dashed line (green) represents only the environment-induced atom-plate interaction.}
\end{figure}

\begin{figure}[htbp]
	\begin{center}	
	    \includegraphics[scale=0.55]{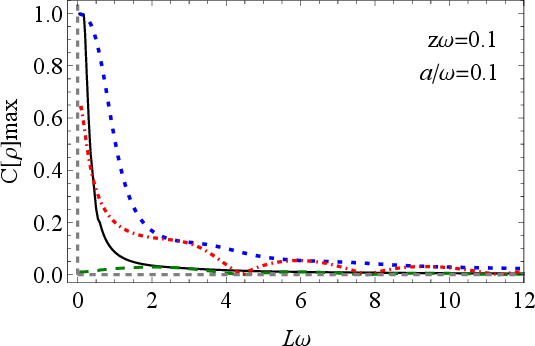}\vspace{0.01\textwidth}\hspace{0.02\textwidth}
		\includegraphics[scale=0.55]{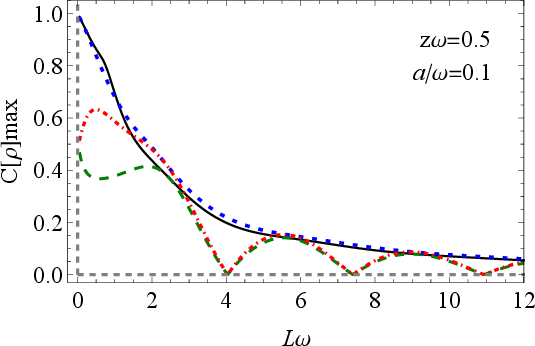}\vspace{0.01\textwidth}\hspace{0.02\textwidth}
		\includegraphics[scale=0.55]{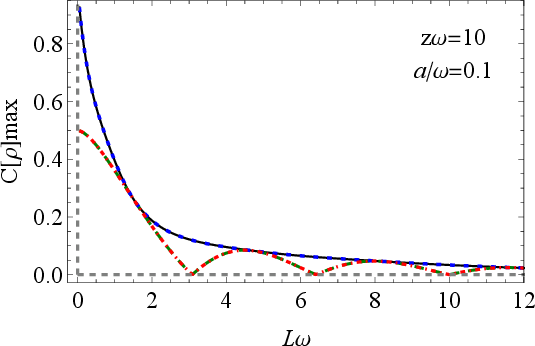}\vspace{0.01\textwidth}
		\includegraphics[scale=0.55]{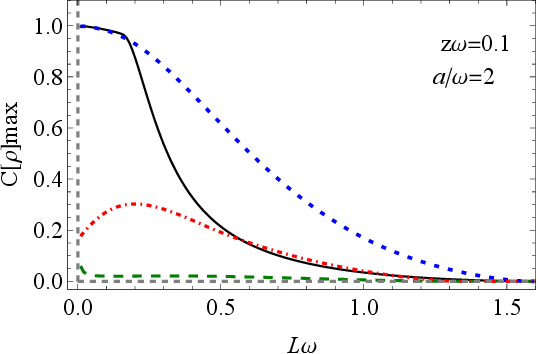}\vspace{0.01\textwidth}\hspace{0.02\textwidth}
		\includegraphics[scale=0.55]{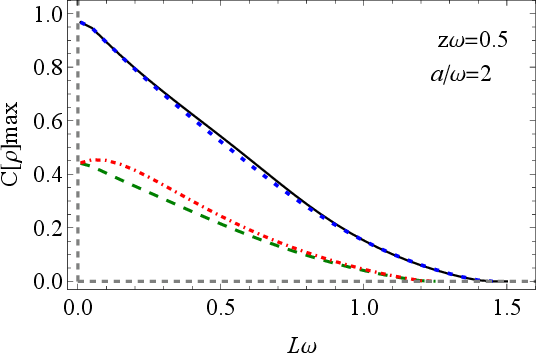}\vspace{0.01\textwidth}\hspace{0.02\textwidth}
		\includegraphics[scale=0.55]{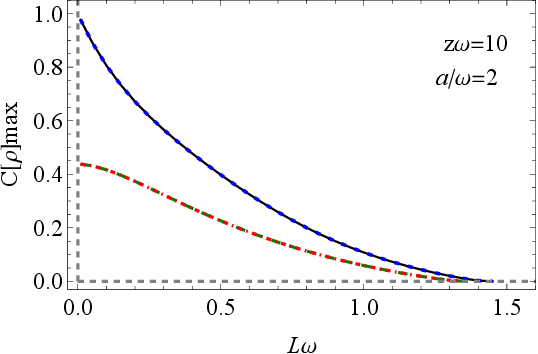}\vspace{0.01\textwidth}
\end{center}\caption{\label{Lwmax} The maximum of concurrence $C[\rho]_{max}$ during evolution is plotted as a function of $L\omega$ with $z\omega=\{0.1, 0.5, 10\}$ and $a/\omega=\{0.1, 2\}$ initially in the state $\left| 10 \right\rangle$. The solid (black) and dot-dashed (red) lines represent uniformly accelerated atoms with and without environment-induced interactions, respectively. The dotted line (blue) represents only the environment-induced atom-atom interaction, and the dashed line (green) represents only the environment-induced atom-plate interaction.}
\end{figure}

We describe the maximum of concurrence during evolution as a function of acceleration in Fig. \ref{awmax}. When we take a small $z\omega=0.1$, for a small $L\omega=0.4$, the environment-induced interatomic interaction dominates, and $C[\rho]_{max}$ is larger compared to the case without interactions. For a not small $L\omega$, since either the environment-induced interatomic interaction or the environment-induced atom-plate interaction dominates, $C[\rho]_{max}$ can be larger or smaller compared to the case without interactions. Taking an intermediate $z\omega=0.5$, when $L\omega<L_{c}\omega$ ($L_{c}\omega\approx1.081$), $C[\rho]_{max}$ is larger with environment-induced interactions considered. For $L\omega>L_{c}\omega$, $C[\rho]_{max}$ may be smaller at a certain small acceleration when the environment-induced interactions are considered. For a large $z\omega=10$, $C[\rho]_{max}$ is dominated by the environment-induced interatomic interaction and exhibits oscillatory behavior for small accelerations.

As shown in Fig. \ref{Lwmax}, we plot the maximum concurrence during evolution as a function of the interatomic separation. For a small $z\omega=0.1$, the environment-induced interatomic interaction dominates for small interatomic separations, and $C[\rho]_{max}$ is larger than in the case without the environment-induced interactions. As $L\omega$ increases, $C[\rho]_{max}$ may be either larger or smaller than in the case without the environment-induced interactions. When we take an intermediate $z\omega=0.5$ and a small $a/\omega=0.1$, $C[\rho]_{max}$ may be smaller than in the case without interactions for certain values of $L\omega$. For a large $a/\omega=2$, the environment-induced interatomic interaction dominates, and $C[\rho]_{max}$ is larger than in the case without interactions. Taking a large $z\omega=10$, the environment-induced atom-plate interaction can be neglected.

\section{Conclusion}
In the Minkowski vacuum, taking the environment-induced interactions into account, we have studied the entanglement dynamics of uniformly accelerated atoms coupled with fluctuating massless scalar fields in the presence of a reflecting boundary. The atoms are aligned perpendicular to the reflecting boundary.

We first consider the two-atom system initially prepared in the state $\left| 10 \right\rangle$. For a small interatomic separation, the environment-induced interactions can be ignored at later times with a large $a/\omega$ or a large $z\omega$. Taking an intermediate interatomic separation, higher peaks of $C[\rho]$ can be obtained for certain parameters with the environment-induced interactions considered. For a large interatomic separation and a small $z\omega$, when $a/\omega$ lies between the two critical values, the environment-induced atom-plate interaction becomes dominant at later times and suppresses entanglement generation, whereas the environment-induced interatomic interaction can be ignored. We then consider the initial state $\left| A \right\rangle$. For a not large $L\omega$ and $z\omega$, with a not small $a/\omega$, the decay of entanglement is very rapid, but the environment-induced interactions have only a small effect on the decay.

For the two-atom system initially prepared in the state $\left| 10 \right\rangle$, we study the maximum of concurrence $C[\rho]_{max}$ generated during evolution as a function of the distance between the nearer atom and the plate. For a not large $L\omega$, $C[\rho]_{max}$ exhibits oscillatory behavior for small $a/\omega$ and approaches a steady value as $z\omega$ increases, but the oscillatory behavior disappears when $a/\omega$ is not small. Taking a large $L\omega$, when the environment-induced interactions are considered, for a small $a/\omega$, the oscillatory behavior of $C[\rho]_{max}$ persists in the large $z\omega$ regime. The maximum concurrence with respect to acceleration has been studied. When we take a small $z\omega$ and $L\omega$, $C[\rho]_{max}$ is larger compared to the case without interactions. For a large $z\omega$, $C[\rho]_{max}$ exhibits oscillatory behavior at small accelerations. The maximum concurrence as a function of the interatomic separation has also been studied. For a small $z\omega$, the environment-induced interatomic interaction dominates for small interatomic separations, and $C[\rho]_{max}$ is larger than in the case without interactions. As $L\omega$ increases, $C[\rho]_{max}$ may be either larger or smaller than in the case without interactions.


\end{document}